\documentclass[aps,prl,twocolumn,superscriptaddress]{revtex4-2}
\usepackage{graphicx}
\usepackage{bm}
\usepackage{xcolor}
\usepackage{amsmath}
\usepackage{amssymb}
\usepackage{hyperref}
\usepackage{tabularray}
\usepackage{wasysym}
\usepackage{ulem}

\begin{document}
	
	\title{Sub-spin-flop switching of a fully compensated antiferromagnet by magnetic field}
	\author{Honglin Zhou}
	\affiliation{Beijing National Laboratory for Condensed Matter Physics, Institute of Physics, Chinese Academy of Sciences, Beijing 100190, China}
	\affiliation{School of Physical Sciences, University of Chinese Academy of Sciences, Beijing 100190, China}
	\author{Muyu Wang}
	\affiliation{Beijing National Laboratory for Condensed Matter Physics, Institute of Physics, Chinese Academy of Sciences, Beijing 100190, China}
	\affiliation{School of Physical Sciences, University of Chinese Academy of Sciences, Beijing 100190, China}
	\author{Yinina Ma}
	\affiliation{Beijing National Laboratory for Condensed Matter Physics, Institute of Physics, Chinese Academy of Sciences, Beijing 100190, China}
	\author{Xiaoyan Ma}
	\affiliation{Beijing National Laboratory for Condensed Matter Physics, Institute of Physics, Chinese Academy of Sciences, Beijing 100190, China}
	\affiliation{School of Physical Sciences, University of Chinese Academy of Sciences, Beijing 100190, China}
	\author{Gang Li}
	\affiliation{Beijing National Laboratory for Condensed Matter Physics, Institute of Physics, Chinese Academy of Sciences, Beijing 100190, China}
	\author{Zihao Tao}
	\affiliation{International Center for Quantum Materials, School of Physics, Peking University, Beijing 100871, China}
	\author{Xiquan Zheng}
	\affiliation{International Center for Quantum Materials, School of Physics, Peking University, Beijing 100871, China}
	\author{Liqin Yan}
	\affiliation{Beijing National Laboratory for Condensed Matter Physics, Institute of Physics, Chinese Academy of Sciences, Beijing 100190, China}
	\affiliation{School of Physical Sciences, University of Chinese Academy of Sciences, Beijing 100190, China}
	\author{Yingying Peng}
	\affiliation{International Center for Quantum Materials, School of Physics, Peking University, Beijing 100871, China}
	\author{Ding-Fu Shao}
	\email{dfshao@issp.ac.cn}
	\affiliation{Key Laboratory of Materials Physics, Institute of Solid State Physics, HFIPS, Chinese Academy of Sciences, Hefei 230031, China}
	\author{Bo Liu}
	\email{liubo@iphy.ac.cn}
	\affiliation{Beijing National Laboratory for Condensed Matter Physics, Institute of Physics, Chinese Academy of Sciences, Beijing 100190, China}
	\author{Shiliang Li}
	\email{slli@iphy.ac.cn}
	\affiliation{Beijing National Laboratory for Condensed Matter Physics, Institute of Physics, Chinese Academy of Sciences, Beijing 100190, China}
	\affiliation{School of Physical Sciences, University of Chinese Academy of Sciences, Beijing 100190, China}
	\begin{abstract}
		
		The control of antiferromagnets by magnetic fields represents a fundamental challenge in condensed matter physics, owing to their fully compensated magnetic order and vanishing net magnetization. Conventional methods rely on either uncompensated moments or high-field spin-flop transitions. Here, we demonstrate low-field switching in the fully compensated antiferromagnet CeNiAsO---a material recently proposed as a candidate for $p$-wave magnetism. Using an in-plane magnetic field well below the spin--flop threshold, we selectively stabilize one of two degenerate antiferromagnetic domains with mutually orthogonal sublattice orientations. This field-induced domain selection allows reversible and nonvolatile switching of a giant in-plane resistivity anisotropy up to $\sim$35\,\%---a magnitude that far exceeds conventional anisotropy signals driven by spin--orbit coupling. The switching behavior persists across both the low-temperature noncollinear N\'{e}el  phase and the higher-temperature collinear spin-density-wave phase, highlighting the universality of the domain-selection mechanism. Our work establishes a practical approach for manipulating compensated antiferromagnets with modest magnetic fields and underscores their potential for high-performance spintronic devices based on giant and switchable resistivity anisotropy.

	\end{abstract}
	
	\maketitle
	The application of magnetic fields constitutes one of the most fundamental and effective means to control the electronic properties of materials. In magnetic systems in particular, realigning magnetic moments with an external magnetic field can dramatically alter electronic transport and lead to phenomena such as magnetoresistance and zero-field anisotropic resistivity. This ability is essential both for probing fundamental magnetic interactions and spin-dependent scattering processes and for enabling key technologies such as magnetic data storage and sensing. While highly effective in ferromagnets and ferrimagnets—owing to their net magnetization—magnetic manipulation of antiferromagnets remains challenging \cite{BaltzV18,JungwirthT18,ChenH24,RimmlerB25}. Their fully compensated magnetic structure yields zero net moment, making antiferromagnets largely “invisible” to moderate magnetic fields and thus difficult to control. 
	
	To date, two primary mechanisms allow magnetic-field control in antiferromagnets. The first requires symmetry breaking that introduces a small net magnetization, enabling indirect reorientation of the  N\'{e}el vector via field-induced torque of the net moment. This approach, however, does not apply to perfectly compensated, high-symmetry antiferromagnets. The second mechanism is the spin-flop effect \cite{Neel_1936,Neel_1952}, in which a sufficiently strong magnetic field applied causes the sublattice moments to reorient abruptly to the plane perpendicular to the field, and then forms a canted antiferromagnetic (AFM) state that evolves with increasing field strength. Although effective, this process typically requires large, often impractical magnetic fields and alters the original AFM ground state. 	As a result, controlling fully compensated antiferromagnets with realistic magnetic fields remains a fundamentally outstanding challenge. 
	
	Compounding this difficulty, the electrical detection of antiferromagnetic order presents an equally formidable obstacle. Conventional probes such as neutron scattering or specialized optical techniques \cite{nvemec2018} demand sophisticated instrumentation and are incompatible with standard device architectures. While spin currents associated with AFM order can be generated, their electrical readout generally relies on engineered heterostructures—such as antiferromagnetic tunnel junctions (exploiting tunneling magnetoresistance) \cite{shao2024AFMTJ}  or antiferromagnet/heavy-metal bilayers (utilizing the inverse spin Hall effect) \cite{Hoogeboom2017}. Direct electrical transport measurements within a single-phase AFM material typically depend on relativistic spin-orbit coupling effects, such as anisotropic magnetoresistance, which usually produce weak signals \cite{wadley2016electrical,bodnar2020magnetoresistance,marti2014room,kriegner2016multiple,PhysRevLett.109.137201,yan2019piezoelectric,feng2025nonvolatile,nair2020electrical,xu2020anisotropic,wang2019giant,duttagupta2020spin}. Even the anomalous Hall effect, allowed only in systems with broken time-reversal and spatial symmetries, typically exhibits a Hall angle on the order of 1\,\%, limiting its utility for practical applications \cite{vsmejkal2022anomalous}. Consequently, achieving both efficient control \emph{and} robust electrical readout of AFM states in a single, fully compensated material has remained elusive. 
	
	In this work, we demonstrate successful manipulation of CeNiAsO \cite{LuoY11,WuS19,LuF23}, a fully compensated antiferromagnet, using a magnetic field well below the spin-flop threshold—thereby overcoming the long-standing control challenge—and simultaneously realize highly efficient electrical detection via a giant, switchable resistivity anisotropy. We report a large twofold in-plane resistivity anisotropy ($\sim$35\,\%) between two perpendicular crystalline axes. By applying an in-plane magnetic field, we achieve reversible and nonvolatile switching of the resistivity anisotropy upon alternating the field direction. We attribute this effect to magnetic-field-induced lifting of degeneracy between two symmetrically equivalent AFM domains with mutually orthogonal sublattice orientations; the field selectively stabilizes one domain, thereby switching the resistivity anisotropy. Remarkably, this switching is observed in both the ground-state noncollinear AFM phase and the higher-temperature collinear spin-density-wave (SDW) phase, suggesting that the underlying domain-selection mechanism is universal across different magnetically ordered states.

	\begin{figure}[tbp]
		\includegraphics[width=\columnwidth]{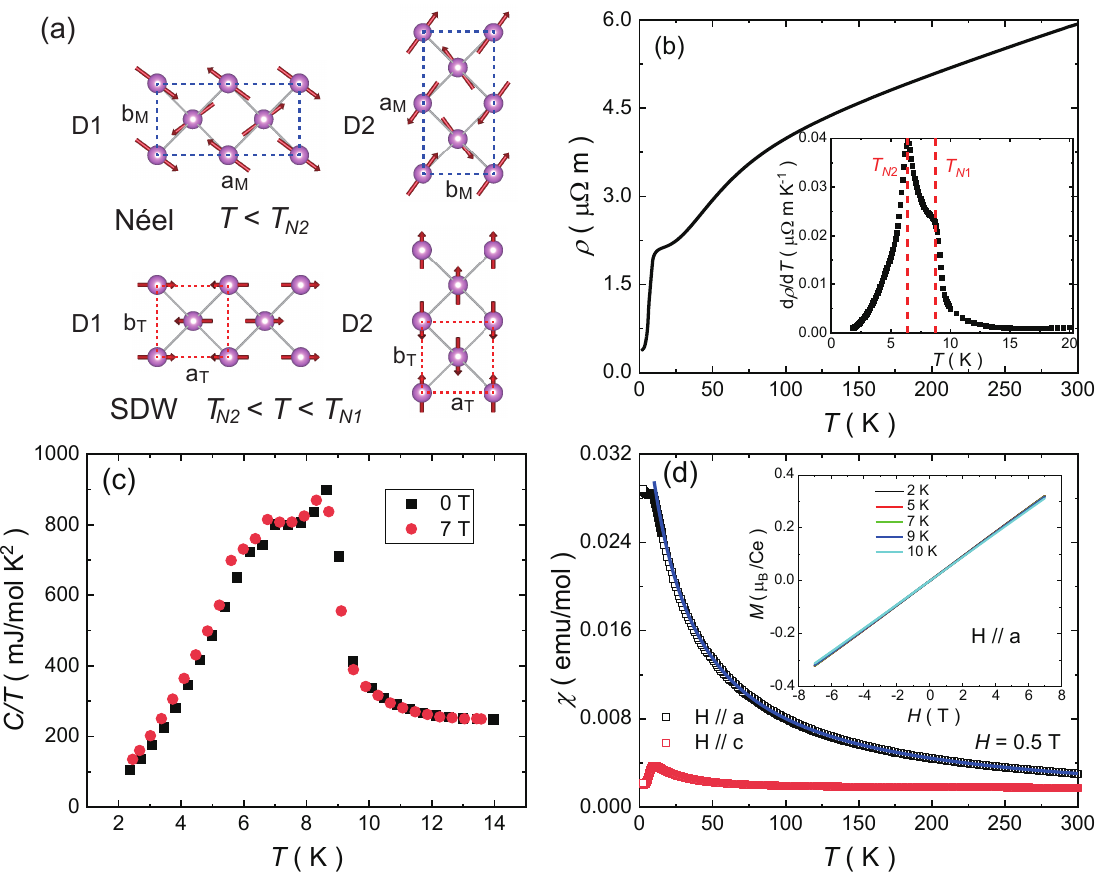}
		\caption{(a) Sketch of the commensurate N\'{e}el (upper panel) and incommensurate SDW (lower panel) in-plane magnetic structures of CeNiAsO for $T < T_{N2}$ and $T_{N2} < T < T_{N1}$, respectively \cite{WuS19}. Only cerium atoms are shown for simplicity. The red and blue dashed lines define the nuclear and  N\'{e}el-order magnetic in-plane unit cells, respectively. Note that the full incommensurate modulation in the SDW state is not displayed since only one magnetic unit cell for the N\'{e}el state is shown.  While $a_T$ and $b_T$ lattice constants are the same in the tetragonal symmetry, lattice constants $a_M$ and $b_M$ in the magnetic unit cell are equal to $2a_T$ and $b_T$, respectively. This results in two domains (D1 and D2) for each AFM order. (b) Temperature dependence of in-plane resistivity. The inset shows its first derivative at low temperatures where $T_{N1}$ and $T_{N2}$ are labeled. (c) The temperature dependence of specific heat $C/T$ below 14 K at 0 and 7 T. (d) Temperature dependence of magnetic susceptibility $\chi=M/H$ for the field $H$ parallel to $a$ and $c$ axes, respectively. The blue line is the Curie-Weiss fit result for the $\chi_{a}$. The inset shows the $M$-$H$ loop at various temperatures with $H//a$. }
		\label{basic}
	\end{figure}
	
	We provide a brief overview of the crystal and magnetic structures of CeNiAsO before presenting our data. CeNiAsO is a structural analogue of the 1111-type iron-based superconductors, crystallizing in the tetragonal ZrCuSiAs-type structure with space group $P4/nmm$ (No. 129)  \cite{LuoY11}. Its lattice constants are comparable to those of cuprates and iron-based superconductors, which may favor coupling with high-temperature superconductivity. This compound undergoes two successive AFM transitions at $T_{N1}$ and $T_{N2}$ with magnetic moments on Ce ions \cite{LuoY11,WuS19,LuF23}. Both pressure and phosphorus doping suppress the AFM orders and induce a quantum critical point \cite{LuoY14}. These results highlight the heavy-fermion behavior of the electronic system. 
	The magnetic ground state of CeNiAsO exhibits a commensurate N\'{e}el order, characterized by a noncollinear magnetic arrangement of moments within the $ab$ plane without $c$-axis canting \cite{WuS19}. This results in a magnetic unit cell that doubles the tetragonal crystallographic unit cell, with in-plane lattice constants satisfying $a_M$ = 2$a_T$ and $b_M$ = $b_T$, where the subscripts $M$ and $T$ denote the magnetic and tetragonal unit cells, respectively [Fig.\ref{basic}(a)]. Specifically, the Ce moments exhibit ferromagnetic alignment along the $b_M$ axis and AFM alignment along the $a_M$ axis, with the moments canted at approximately 36$^\circ$ relative to the $a_M$ axis. This magnetic structure gives rise to two energetically degenerate domains, denoted as D1 and D2. These domains are related by a $C_4$ rotation about the $c$-axis [Fig.\ref{basic}(a)]. Between $T_{\mathrm{N2}}$ and $T_{\mathrm{N1}}$, the N\'{e}el-ordered states in both D1 and D2 evolve into a collinear incommensurate spin density wave (SDW) phase. In this phase, stripe-like ferromagnetic sublattices form along the $b_M$-direction, with the N\'{e}el vector oriented along the $a_M$-direction [Fig.\ref{basic}(a)]. Above $T_{\mathrm{N1}}$, CeNiAsO enters the paramagnetic phase.
	Recently, CeNiAsO has been proposed as a candidate to host $p$-wave magnetism with many intereting phemonema in its noncollinear AFM ground state \cite{HellenesAB23,MaedaK24,ChakrabortyA25,SukhachovP24,EzawaM24,NagaeY25,KokkelerT25,FuyakaY25,SukhachovP25,SooriA25,SalehiM25}.
	
	The presence of these magnetic phases is clearly revealed in our transport measurements using the CeNiAsO single crystals grown via a two-step process \cite{supp}. Figure \ref{basic}(b) displays the temperature dependence of in-plane resistivity at 0 T, with its first derivative shown in the inset. $T_{N1}$ and $T_{N2}$ are 8.8  and 6.2 K, respectively. Figure \ref{basic}(c) presents the low-temperature specific heat versus temperature, where a 9-T field has almost no effect on the specific heat. These results confirm two AFM transitions consistent with prior polycrystalline studies \cite{LuoY11}. Figure \ref{basic}(d) shows temperature-dependent magnetic susceptibility, revealing strong magnetic anisotropy for $H//a$ and $H//c$ from 300 K down to 2 K. The high-temperature $H//a$ data ($\chi_a$) can be fitted by the Curie-Weiss function, $\chi = \chi_0 + C/(T-\theta_W)$, where $\chi_0$, $C$, and $\theta_W$ are the temperature-independent background, Curie constant, and Weiss temperature, respectively. The fitting yields $\theta_W$ = -23.8 K and $\mu_{eff}$ = 2.11 $\mu_B$, close to those reported for polycrystalline samples \cite{LuoY11,WuS19}. The weak temperature dependence of $\chi_{c}$ prevents reliable Curie-Weiss fitting.
	
	\begin{figure}[tbp]
		\includegraphics[width=\columnwidth]{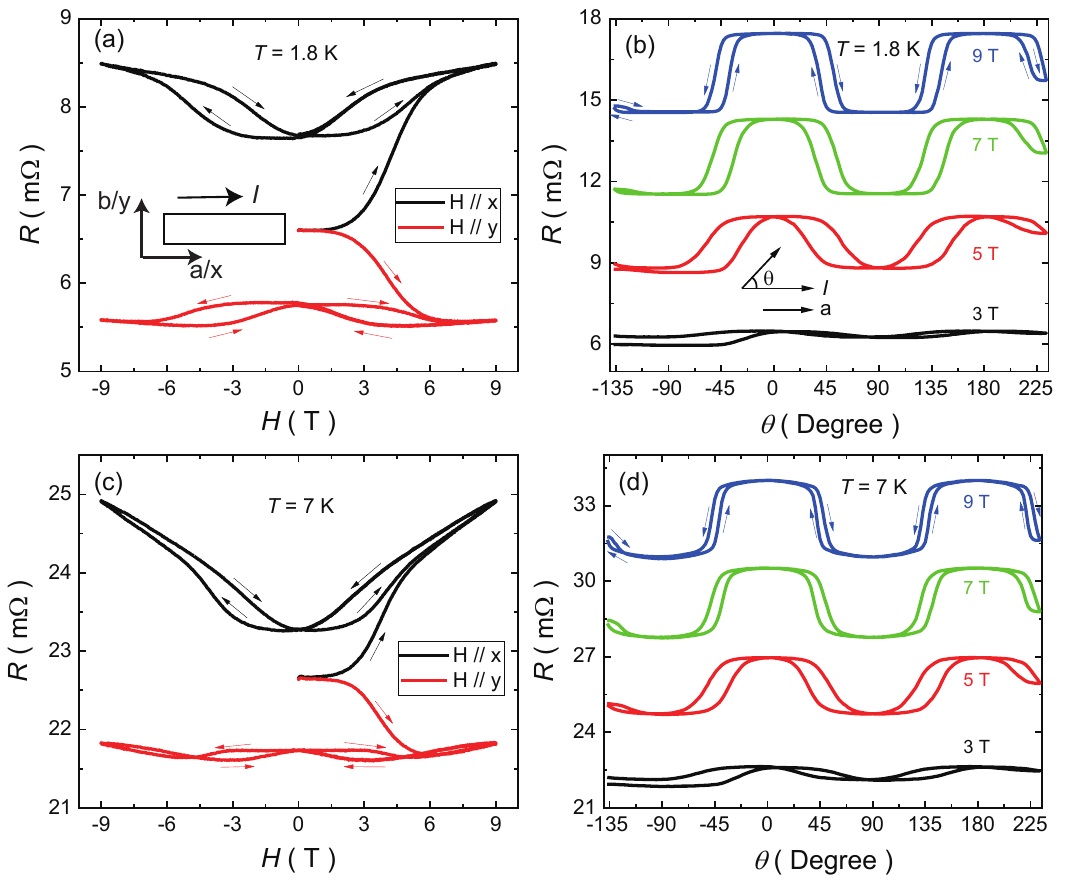}
		\caption{(a) Field dependence of resistance at 1.8 K. The current is applied along the lattice $a$-axis. $x$ and $y$ directions are defined as parallel and perpendicular to the current direction, respectively. The arrows indicate the field application process. (b) Angular dependence of resistance at 1.8 K. The arrows indicate ascending or descending angle scans. (c) Field dependence of resistance at 7 K. (d) Angular dependence of resistance at 7 K. The resistance values in (b) and (d) have been shifted by a constant at each field for better presentation.}
		\label{MR}
	\end{figure}
	
	To measure the in-plane resistivity anisotropy, we apply an in-plane magnetic field---along either the $x$- or $y$-axis---to lift the degeneracy between D1 and D2 by introducing an energy splitting between the two domains. Here $x$- and $y$-axis is defined as the field parallel and vertical to the current flowing along the lattice $a$ axis, respectively [Inset of Fig. \ref{MR}(a)]. Figure \ref{MR}(a) shows the field dependence of resistance at 1.8 K. For $H // x$, the resistance increases rapidly from the ZFC value [$R_{ZFC}$(0T)] with increasing field without saturation up to 9 T. Afterwards, the sample remains in a high-resistance state. Intriguingly, butterfly-shaped hysteresis emerges, and the field-trained zero-field resistance [$R_{FT}$(0T)] (i.e., zero-field resistance after field sweeping up to 9 T and then down to 0 T) does not depend on field application history anymore.  A complementary behavior is observed for $H//y$, where the resistance initially decreases with increasing field and then the sample stays in a low-resistance state. 
	
	To better understand the anisotropic resistance, we measured the angular-dependent magnetoresistance (AMR) with the field rotating within the $ab$ plane. Figure \ref{MR}(b) shows the results at 1.8 K. All curves exhibit 180-degree periodicity, confirming the in-plane two-fold symmetry of resistivity anisotropy. At 9 T, the AMR follows a nearly squared functional form rather than trigonometric functions, indicating that the anisotropy is not associated with the projection of the magnetic field. Significant hysteresis emerges between ascending and descending angular sweeps at all fields. We also perform magnetoresistance (MR) and angular magnetoresistance (AMR) measurements at various temperatures. Interestingly, similar behaviors are observed at 7\,K [Figs.~\ref{MR}(c) and \ref{MR}(d)], where the sample resides in the incommensurate spin-density wave (SDW) state, albeit with minor differences in the hysteresis features. These results suggest that the sample exhibits strong in-plane resistivity anisotropy with two-fold symmetry in both the N\'{e}el and the SDW states. 
	
	\begin{figure}[tbp]
		\includegraphics[width=\columnwidth]{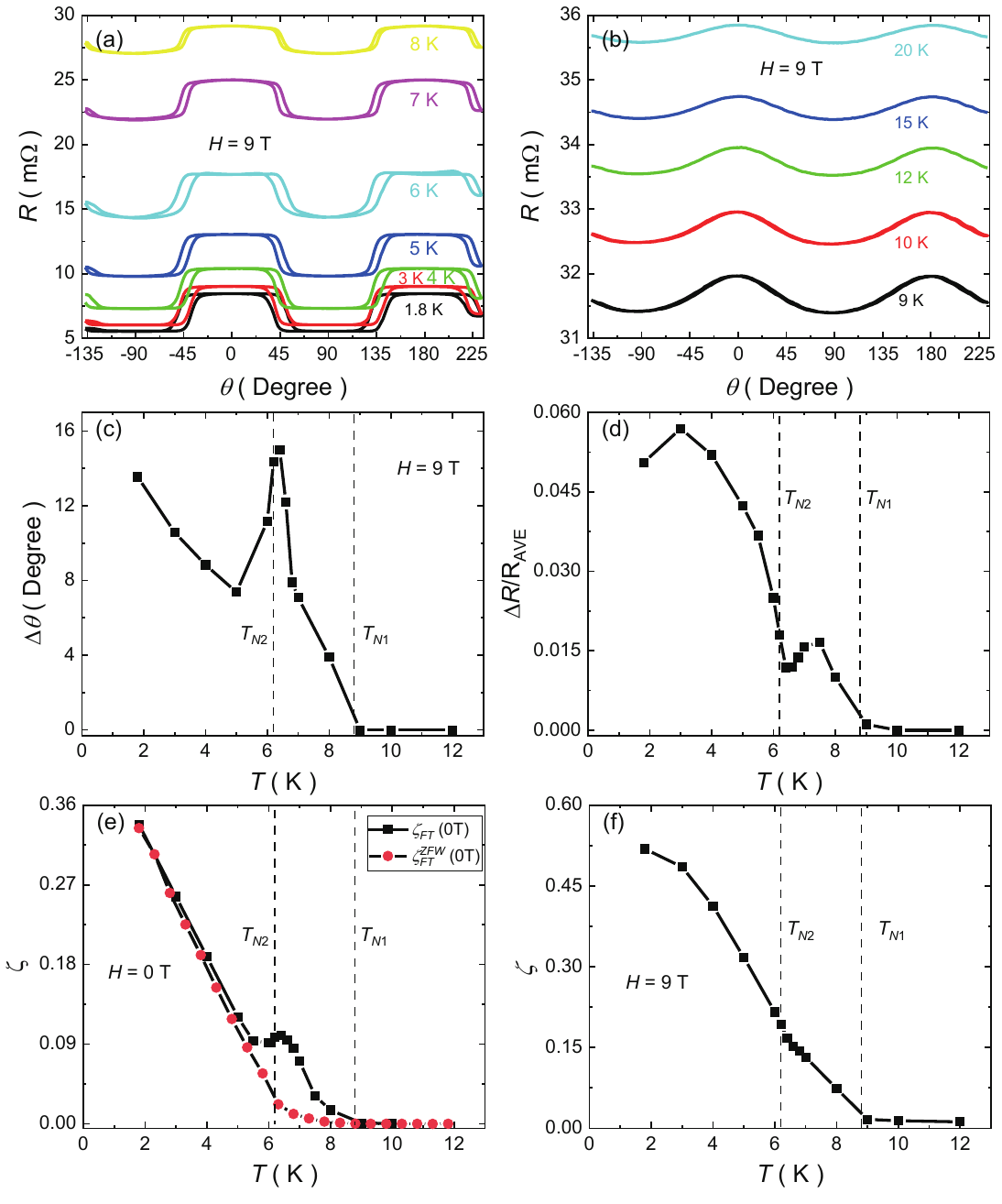}
		\caption{(a) and (b) Angular dependence of resistance at 9 T below and above $T_{N1}$, respectively. (c) and (d) Temperature dependence of the hysteretic angle $\delta \theta$ in AMR at 9 T and maximum hysteretic resistance $\Delta R$ in MR, respectively. (e) Temperature dependence of anisotropic ratio $\zeta$(0T) (as defined in the text). Both $\zeta_{FT}$(0T) and $\zeta_{FT}^{ZFW}$(0T) are measured at zero field. However, the former is measured after 9-T field training at each temperature, while the latter is obtained after 9-T field training at 1.8 K followed by zero-field warming. (f) Temperature dependence of $\zeta$(9T).}
		\label{AMR}
	\end{figure}
	
	Figures \ref{AMR}(a) and \ref{AMR}(b) display the AMR at 9 T below and above $T_{N1}$, respectively. The squared two-fold angular dependence persists up to 8 K and vanishes near $T_{N1}$. Well above $T_{N1}$, the AMR follows a conventional trigonometric form that originates from the Lorentz effect. Although the hysteresis behavior shows no significant difference in different phases, $\Delta\theta$ at 9 T, defined as the angular difference between ascending and descending angular scans, decreases initially with increasing temperature, peaks at $T_{N2}$ and decreases rapidly to zero approaching $T_{N1}$ [Fig. \ref{AMR}(c)]. Intriguingly, while $\Delta R/R_{AVE}$, defined as the ratio of the difference between maximum and minimum resistance at the largest MR hysteresis to their average value, shows a similar temperature trend, its peak lies between $T_{N1}$ and $T_{N2}$ with a minimum at $T_{N2}$ [Fig. \ref{AMR}(d)].

	To quantitatively study the resistivity anisotropy, we define the dimensionless anisotropic ratio $\zeta$ as $(R_{x}-R_{y})/R_{y}$, where the subscripts denote the field orientation parallel to $x$ and $y$ axis, respectively. Due to hysteresis in MR, $\zeta$ clearly depends on field-application history. To obtain $\zeta$ at zero field, we employ the following field-training process: the sample is zero-field cooled to the target temperature, then subjected to 9 T, and finally returned to zero field. Figure \ref{AMR}(e) shows the temperature dependence of $\zeta_{FT}$(0T), which decreases initially with increasing temperature, exhibits a hump slightly above $T_{N2}$, and approaches zero above $T_{N1}$. $\zeta_{FT}^{ZFW}$(0T), which is field-trained at 1.8 K followed by zero-field warming, shows similar values at low temperatures but exhibits no hump around $T_{N2}$ and decreases rapidly to zero when approaching $T_{N1}$ [Fig. \ref{AMR}(e)]. While $T_{N2}$ is discernible in $\zeta$(0T), it becomes nearly indistinguishable in $\zeta$(9T) as shown in Fig. \ref{AMR}(f). These complicated temperature-dependent behaviors could come from factors including the  relaxation time, the band splitting, the magnetic structure and the ordered moment values \cite{WuS19}, and the pinning strength of magnetic domains.

	\begin{figure}[tbp]
		\includegraphics[width=\columnwidth]{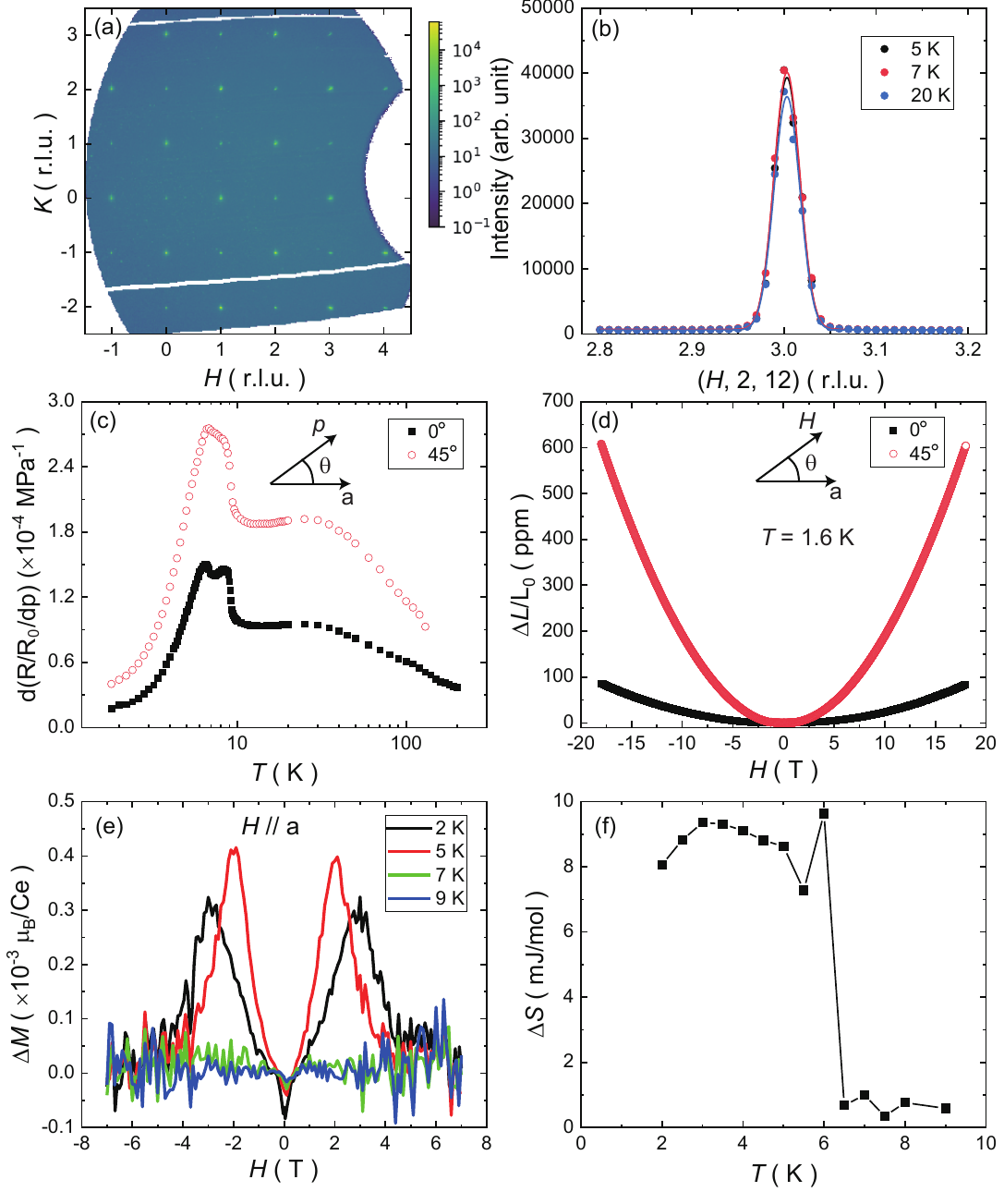}
		\caption{(a) A single-crystal XRD pattern at $L=12$ and $T$ =  5 K. (b) A cut along the $H$ direction for the (3,2,12) Bragg peak at 5, 7, and 20 K. (c) Temperature dependence of elatoresistance $d(R/R_0)/dp$. The inset illustrates the direction of the uniaxial pressure applied. (d) Field dependence of the relative length change at 1.6 K. The inset illustrates the direction of the magnetic field applied. (e) Magnetization difference between decreasing and increasing fields at various temperature. (f) Temperature dependence of the hysteresis loss. }
		\label{origin}
	\end{figure}

	The observed twofold zero-field in-plane resistivity anisotropy suggests a breaking of rotational symmetry from $C_4$ to $C_2$ in the AFM states. As a structural analogue of the 1111-type iron-based superconductors, a natural explanation would involve nematic order, which is commonly accompanied by a tetragonal-to-orthorhombic structural transition and widely observed in iron-based superconductors \cite{FernandesRM14,DaiP15}.
	However, our low-temperature structural analysis of CeNiAsO reveals no structural change from room temperature down to 5 K, as evidenced by single-crystal XRD pattern and the absence of splitting or even broadening in the nuclear Bragg peaks [Figs. \ref{origin}(a) and \ref{origin}(b)]. Furthermore, although the elastoresistance responds to the AFM transitions [Fig. \ref{origin}(c)], its magnitude is significantly weaker than that in iron-based superconductors. Moreover, the temperature dependence of the elastoresistance shows no divergent behavior upon approaching the transitions --- a hallmark of a nematic transition \cite{ChuJH12,LiuZ16,GuY17}.
	Notably, the absence of a structural change also makes the existence of orbital order unlikely, although further investigation is still needed. The magnetorestriction of CeNiAsO exhibits quadratic field dependence up to 18 T yet shows no hysteresis [Fig. \ref{origin}(d)], which also excludes magnetoelastic coupling as the origin of the resistivity anisotropy.
	
	As described above, resistivity anisotropy observed in AFM systems is commonly attributed to AFM domains \cite{PhysRevLett.109.137201,feng2025nonvolatile,nair2020electrical,xu2020anisotropic,wang2019giant,duttagupta2020spin}. Figure \ref{origin}(e) shows the difference in magnetization between decreasing and increasing fields at various temperatures, revealing hysteresis behavior below $T_{N1}$. This is further illustrated by the temperature dependence of the hysteresis loss (i.e., the area of the hysteresis loop) in Fig. \ref{origin}(f). The hysteresis can be attributed to the movement of noncollinear AFM  domains in the N\'{e}el state with anisotropic susceptibility. The absence of hysteresis between $T_{N1}$ and $T_{N2}$ is plausible due to the much weaker anisotropic susceptibility in the SDW state. Notably, magnetic-field control of AFM domains and the resulting resistivity anisotropy have also been observed in iron-based superconductors \cite{ChuJH10}, which is attributed to the nematic order that couples to orthorhombic structure. Accordingly, the magnetoresistance behavior can be well attributed to the remodulattion of the domain populations or alignments. In the ZFC process, the AFM domains distributed randomly due to the tetragonal lattice symmetry and thus the resistivity anisotropy is barely observable. With changing field at low temperatures, one of the domains becomes dominate [Fig. \ref{basic}(a)]. The hysteresis behavior comes from remodulation of the relative concentrations of two domains. This is different from the typical anisotropic magnetoresistance, which comes from the moment rotation under magnetic field due to spin-orbital coupling. 
	
	While the anisotropic resistivity behavior is coupled with magnetic domains, the reason behind this unprecedentedly large anisotropy is worth further discussing. CeNiAsO has been theoretically proposed as a $p$-wave magnet due to the [$C_{2\perp}||t$] spin symmetry of its noncollinear, coplanar magnetic structure in the N\'{e}el state \cite{HellenesAB23}. Accordingly, a large in-plane anisotropic conductivity is predicted. The anisotropic value $(\sigma_{yy}-\sigma_{xx})/(\sigma_{xx}+\sigma_{yy})$, as defined in Ref. \cite{HellenesAB23} (Note that the definition is different from $\zeta$), is approximately 0.15 at 1.8 K in our measurements [Fig. \ref{MR}(a)], which quantitatively agrees with the theoretically calculated value. While this seems to support for $p$-wave magnetism in CeNiAsO, the similar anisotropic behavior in the SDW state [Fig. \ref{MR}(c)] casts a shadow over such an exotic explanation. The pronounced structural anisotropy of the stripe-like magnetic sublattice may provide an alternative mechanism for the observed phenomenon, potentially enabling N\'{e}el spin currents along the stripe direction while suppressing them perpendicular to it \cite{shao2023}. Further theoretical and experimental studies are therefore needed to unambiguously determine the microscopic origin of the anisotropic resistivity in CeNiAsO.

	\begin{figure}[tbp]
		\includegraphics[width=\columnwidth]{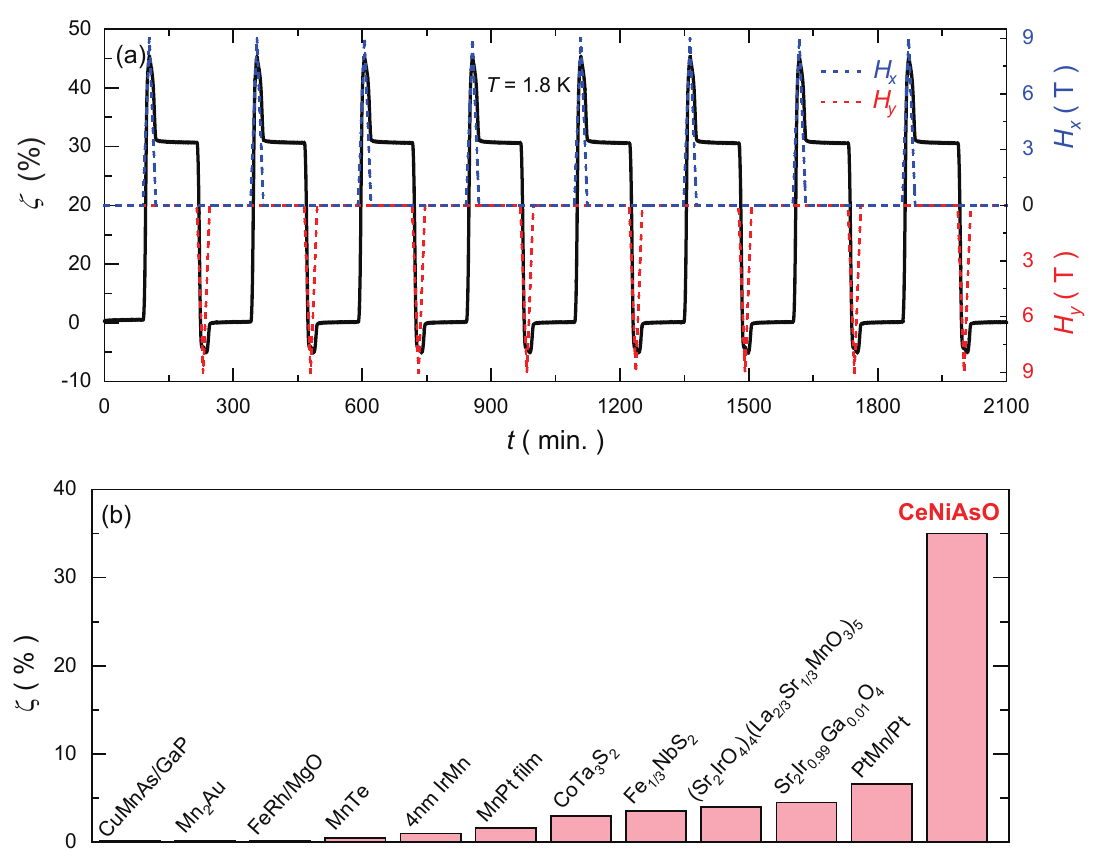}
		\caption{ (a) Control of $\zeta$(0T) at 1.8 K with magnetic field alternating along $x$ and $y$ directions. (b) $\zeta$ values in various AFM systems \cite{wadley2016electrical,bodnar2020magnetoresistance,marti2014room,kriegner2016multiple,PhysRevLett.109.137201,yan2019piezoelectric,feng2025nonvolatile,nair2020electrical,xu2020anisotropic,wang2019giant,duttagupta2020spin}.}
		\label{sum}
	\end{figure}

	Our results demonstrates a nonvolatile, field-driven domain selection and resistivity switching in CeNiAsO and its potential for field-controlled memories. We therefore implement a writing protocol using pulse-like in-plane magnetic fields to deterministically train the domains. The written states are then read out by exploiting the resistance states associated with the domain configuration [Fig. \ref{sum}(a)]. At 1.8 K, reversible switching between high- and low-resistance states is achieved by applying alternating magnetic field pulses along the x- and y-directions. This write-read cycle can be repeated over multiple iterations without degradation in resistance contrast, demonstrating robustness and non-volatility. The resistivity anisotropy in the AFM state of CeNiAsO is remarkably large compared to that in other AFMs. As shown in Fig. \ref{sum}(b), the zero-field $\zeta$ for CeNiAsO is nearly an order of magnitude greater than that of other AFM systems \cite{wadley2016electrical,bodnar2020magnetoresistance,marti2014room,kriegner2016multiple,PhysRevLett.109.137201,yan2019piezoelectric,feng2025nonvolatile,nair2020electrical,xu2020anisotropic,wang2019giant,duttagupta2020spin}.
	
	These results have significant implications for AFM spintronics: It is widely believed that controlling antiferromagnets with an external magnetic field is extremely difficult below the spin-flop transition. However, our work demonstrates that the magnetic field can effectively manipulate the relative populations of magnetic domains, leading to a pronounced and nonvolatile resistance change. The observed anisotropy ($>$ 30\%) implies a large On/Off ratio comparable to that of conventional spintronic devices based on the giant magnetoresistance (GMR) effect --- yet achieved without the need for complex multilayer structures.

	In summary, we have demonstrated a dual breakthrough in the manipulation and electrical readout of a fully compensated antiferromagnet. By applying an in-plane magnetic field well below the spin-flop threshold, we achieve nonvolatile, reversible switching of the N\'{e}el order between two orthogonal domains in CeNiAsO. Crucially, this switching is directly manifested as a giant in-plane resistivity anisotropy up to $\sim$35\,\%, which far exceeds conventional spin--orbit-coupling-driven signals and enables robust, all-electrical detection using standard transport measurements in a single-phase material. The effect persists across distinct magnetic phases—both the low-temperature noncollinear AFM state and the higher-temperature collinear SDW phase—highlighting the universality of the underlying domain-selection mechanism.  Our work establishes a practical pathway to control and probe compensated antiferromagnets with modest magnetic fields and simple device geometries, opening new opportunities for antiferromagnetic spintronics based on intrinsic, large-magnitude transport signatures.
	
	S. L. acknowledges helpful discussions with Prof. Quansheng Wu and Prof. Guoqiang Yu. This work is supported by the National Key Research and Development Program of China (Grants No. 2022YFA1403400, No. 2021YFA1400400, No. 2022YFA1403800), the National Natural Science Foundation of China (Grants Nos. 12274411, 12241405, 52250418, and 12374143), the Basic Research Program of the Chinese Academy of Sciences Based on Major Scientific Infrastructures (Grant No. JZHKYPT-2021-08), and the CAS Project for Young Scientists in Basic Research (Grant No. YSBR-084). This study was also partially supported by  Synergetic Extreme Condition User Facility (SECUF).

\end{document}